 \pdfoutput=1 
\documentclass[reprint,amsmath,amssymb,aps,]{revtex4-2}
\usepackage{graphicx}
\usepackage{graphics}
\usepackage{latexsym}
\usepackage{amsmath, amsthm, amssymb, wasysym}
\usepackage{xcolor}
\usepackage{epsfig}
\usepackage{soul}
\usepackage{cancel}
\usepackage{hyperref}

\begin{document}
\setstcolor{red}
\title{ The Emergence of Laplace Universality in Correlated Processes 
}

\author{R. K. Singh$^{1,2}$}
\author{Stanislav Burov$^{1}$}
\affiliation{$^1$Department of Physics, Bar-Ilan University, Ramat-Gan 5290002,
Israel}
\email{stasbur@gmail.com}
\affiliation{$^2$Department of Biomedical Engineering, Ben-Gurion University of the Negev, Be’er Sheva 85105, Israel}



\begin{abstract}
In transport processes across materials like glasses, living cells, and porous media, the probability density function of displacements exhibits exponential decay rather than Gaussian behavior. We show that this universal behavior of rare events, termed Laplace tails, emerges even when correlations and memory affect the dynamics.
Using a renormalization-based approach, we demonstrate that correlations and memory do not suppress these tails but rather enhance their visibility, even at short timescales. 
The developed analytical framework refines the concept of correlations for rare events and enables the computation of effective parameters that govern Laplace tails in correlated processes.
These findings suggest that correlations can serve as a tunable parameter to control the behavior of rare events in transport.
\end{abstract}

\maketitle

\newcommand{\fr}{\frac}
\newcommand{\lr}{\langle}
\newcommand{\rl}{\rangle}
\newcommand{\tl}{\tilde}

The ubiquity of Gaussian statistics in many systems stems from the Central Limit Theorem (CLT), which states that the sum of independent and identically distributed (IID) random variables (RVs) tends toward a normal distribution.
Examples include small displacements of a single molecule, which add up to its final position, or daily stock price fluctuations, which add up to the final price.
However, recent experiments revealed the existence of another universality that holds for rare events, known as Laplace universality~\cite{hamdi2024laplace}. 
The probability density function (PDF) of the position of transport carriers in heterogeneous materials has been observed to exhibit an exponential decay for large displacements~\cite{chaudhuri2007universal,wang2009anomalous,wang2012brownian}, termed Laplace tails.
This phenomenon has been documented in numerous systems, including glass-forming liquids~\cite{rusciano2022fickian}, nanoparticles in polymer melts~\cite{xue2016probing}, colloidal beads~\cite{pastore2021rapid}, polymeric glassformers~\cite{srinivasan2024nature}, living cells~\cite{aaberg2021glass,schramma2023chloroplasts},and insulin granules~\cite{yi2024distinct} among others~\cite{kumar2023anomalous,zu2021emergent,Hapca2009Anomalous,Xue2020Diffusion,Witzel2019Heterogeneities,Cherstvy2019Non,chechkin2017brownian}, stimulating the development of diffusing-diffusivity models~\cite{chubynsky2014diffusing,hidalgo2021cusp,nampoothiri2021polymers,yamamoto2021universal,sposini2018first,sposini2024being}.
It has been shown~\cite{barkai2020packets} that an analogous result to the CLT holds for sums of IID RVs where the number of RVs is the random number of displacements recorded during measurement time $t$. Specifically, the PDF of $X=\sum_{i=1}^{N_t}x_i$ exhibits Laplace tails for large $X$, up to logarithmic corrections~\cite{barkai2020packets}, when $x_i$ are IID, and the number of events $N_t$ follows a renewal process.
In practice, however, the assumption of independence among RVs is rarely met.
This applies to both the Gaussian center and Laplace tails. 
Fluctuations in financial markets~\cite{bouchaud2003fluctuations,bouchaud2006random}, animal movements~\cite{maye2007order,johnson2008continuous}, transport in porous media~\cite{liu2020correlated,le2011effective}, and living cells~\cite{tabei2013intracellular} are examples where memory and correlations play a significant role.
Nevertheless, Gaussian center and Laplace tails are widely observed. 
While Gaussian universality emerges at long timescales, Laplace tails appear even for short times when systemic details control transport and correlations are significant. 
We aim to explain why Laplace tails arise despite correlations and are present even at very short timescales~\cite{Dapeng2023triggering}.  

As stated above, Gaussian statistics emerge even for systems where the single RVs, i.e., displacements, are correlated. 
For example, the short-time displacements of {\it E. Coli} bacteria are correlated, but its long-time position distribution follows Gaussian statistics~\cite{berg1993random}. Similarly, the positions of monomers in a polymer under tension are correlated due to short-range interactions, yet the long-range properties are Gaussian~\cite{de1979scaling}.

The resolution to this apparent contradiction lies in the large number of constituents in the sum and the correlation-reducing properties of coarse-graining.
Any sum of RVs $\sum_{i=1}^N x_i$ can be rewritten in terms of coarse-grained RVs, also known as block RVs.
For example a coarse-graining of the form $(x_{2j}+x_{2j-1})/\sqrt{2}\to{\tilde x}_j$ leaves the sum unchanged, $\sum_{i=1}^N x_i \to \sqrt{2}\sum_{j=1}^{N/2} {\tilde x}_j$. 
If the correlation between the original $x_i$ and $x_j$ is decaying sufficiently fast as a function of $|i-j|$, after several rounds of coarse-grainings, the block variables will be effectively uncorrelated and can be treated as IID. 
Since the sum of these block variables is proportional to the original sum $\sum_{i=1}^N x_i$, the Gaussian statistics will show itself even when $x_i$ are correlated. Note that the Gaussian statistics will be determined by intrinsic properties, like the mean and variance, of the block variables and not the original $x_i$.
Therefore, repeated coarse-graining acts as a renormalization-group flow~\cite{cardy1996scaling}, leading to a Gaussian fixed point for any observable $X=\sum_{i=1}^N x_i$. This holds if the correlation between different $x_i$ has a finite scale and $N$ is large enough for sufficient coarse-graining repetitions to take place. 


We build on the concept of convergence to a functional fixed point through coarse-graining, which accounts for the emergence of the Gaussian center, and extend it to study large deviations in a correlated process $X = \sum_{i=1}^{N_t} x_i$. 
By employing subordination of large deviations, we find that for large $|X|$, the Laplace tails acts as a functional fixed point. 
In contrast to the Gaussian case, the fixed point associated with Laplace tails governs the statistics even at very short measurement times.
Furthermore, while correlations typically modify the Gaussian center at short times, they can significantly enhance the Laplace tails. 
We develop mathematical forms and concepts that replace the traditional correlation function, which relies on average behavior, with a framework suited to addressing rare events.

First, we present the framework that leads to exponentially decaying tails of the PDF. 
The continuous time random walk (CTRW)~\cite{bouchaud1990anomalous,metzler2000random} mimics the hop dynamics tracers exhibited in many systems that display exponential decay. 
The random times between the hops present in CTRW are the key elements behind the Laplace tails~\cite{chaudhuri2007universal,barkai2020packets}. 
 Initially developed to model transport in amorphous materials, the CTRW has proven highly relevant to a wide range of systems~\cite{metzler2000random,kutner2017continuous}. 

The CTRW process is defined as follows: a particle starts at the origin and jumps to a new position after a random time $\tau_1$, then waits for another random time $\tau_2$ before making the next jump, continuing this pattern.
The process terminates when the total elapsed time exceeds the measurement time $t$. The jumps $x_i$ are drawn from a distribution $f(x_i)$, and the waiting times $\tau_i$ from a distribution $\psi(\tau_i)$. The number of jumps $N_t$ by time $t$ is independent of the spatial position and is determined by $\psi(\tau)$.  
The PDF for performing $N$ jumps during time $t$ is given by $Q_t(N)$. 
For a given $N_t$, the spatial position $X = \sum_{i=1}^{N_t} x_i$ is determined by $f(x)$, and the PDF of being at $X$ after $N$ jumps is $P_N(X)$. 
The PDF of finding the process at position $X$ at time $t$, $P(X,t)$, is obtained by summing over all possible values of $N_t = N$ using the law of total probability, a technique known as subordination~\cite{bouchaud1990anomalous}:
\begin{equation}
\label{eq: pxt}
P(X,t) = \sum^\infty_{N=0}P_N(X)Q_t(N)
\end{equation}
Since each waiting time is generated independently of the previous waiting times, the temporal process is a renewal process and the form of $Q_t(N)$ is known in Laplace space~\cite{CoxBook}. 
The distribution $P_N(X)$ describes the sum of $N$ IID RVs and is known in Fourier space. 
All properties of uncorrelated CTRW, including the Gaussian center and Laplace tails, can be derived from Eq.~\eqref{eq: pxt}.

Laplace tails are present when the following conditions are met: (a) $f(x)$ decays faster than an exponential for large $x$, i.e., $\exp(x) f(x){\to} 0$ when $|x|\to\infty$, for example  $f(x) \sim \exp\left[-\left(|x|/\delta\right)^\beta\right]$ with $\beta > 1$, and (b) $\psi(\tau)$ attains a power-series expansion near $\tau = 0$, i.e., $\psi(\tau) \sim C_0 \tau^A + C_1 \tau^{A+1}$ with $A \geq 0$. 
Specifically, in the limit of large $|X|$,
\begin{equation}
\label{eq:pxtLaplaceorigin}
P(X,t)\underset{|X|\to\infty}{\sim}\exp\left\{-\frac{|X|}{\delta}Z\left(\frac{|X|}{t}\right) + t\frac{C_1}{C_0}\right\},
\end{equation}
where $Z(y)$ is a slowly varying function of $y$ introducing logarithmic corrections~\cite{barkai2020packets,wang2020large,pacheco2021large}. $Z(y)$ depends on $\beta$, $\delta$, $A$, and the coefficient $C_0$~\cite{barkai2020packets}. 
The case of $\beta = 1$ represents the critical point between the universal exponential decay described by Eq.~\eqref{eq:pxtLaplaceorigin} and a specific decay following the large $|x|$ behavior of $f(x)$~\cite{PhysRevE.108.L052102}.

The universality stems from the fact that for $\beta>1$ and large $|X|$, the sum in Eq.~\eqref{eq: pxt} is controlled by a dominant term that occurs for a specific $N=N^*$~\cite{barkai2020packets}. 
The scaling of $N^*$ depends on the ratio $\kappa=\frac{|X|}{t}\big/\left(\delta {C_0}^{\frac{1}{A+1}}\right)$. 
When $\kappa \gg 1$, $N^*\sim |X|$ gives rise to the exponential form in Eq.~\eqref{eq:pxtLaplaceorigin}. 
Conversely, as $\kappa \to 0$, $N^*$ follows $N^*\sim t$, which gives rise to the Gaussian center.  For both cases, the universality is achieved due to a large number of jumps, i.e., $N^*$, which can be achieved not only when the elapsed time is large (Gaussian universality) but also when the process has to reach a distant position (Laplace universality).

Laplace tails were proven to arise when all $x_i$ are IID RVs and $\tau_i$ form a renewal process~\cite{barkai2020packets,wang2020large,pacheco2021large}. 
Under these conditions, the Laplace form in Eq.~\eqref{eq:pxtLaplaceorigin} is expected when~\cite{barkai2020packets} 
\begin{equation} 
\label{eq:starttail}
\frac{|X|}{t} > \frac{\left(C_0 \Gamma[A+1]\right)^{\frac{1}{A+1}}}{\left(\beta(\beta-1)\right)^{\frac{1}{\beta}}\left(A+1\right)^{1-\frac{1}{\beta}}} \delta \end{equation} 
The parameters $\delta$, $C_0$, $\beta$, and $A$ define the exponential decay rate of $P(X,t)$, the range of $X$ and $t$, and how frequently rare events, described by Laplace tails, occur. 
Adjusting these parameters affects the appearance or disappearance of Laplace tails in experiments, determining the relevant space and time scales~\cite{Dapeng2023triggering}.


To extend the results to correlated processes~\cite{magdziarz2012correlated,chechkin2009continuous,tejedor2010anomalous,johnson2008continuous,liu2020correlated,schulz2013correlated} , we apply the renormalization group approach and coarse-graining~\cite{cardy1996scaling}. 
In the Ising model with interactions, coarse-graining groups spins into increasingly larger blocks, eventually making the block variables independent. When interactions are not too extended, the system behaves like a free Ising model with renormalized parameters. 
We generalize this approach to the series $x_1, x_2, \dots, x_N$ and $\tau_1, \tau_2, \dots, \tau_N$, by coarse-graining: $x_1 + x_2 \to x^{(1)}_1$, $x_3 + x_4 \to x^{(1)}_2$, and so on, with similar transformations for $\tau_i$s. After $k$ iterations, the block variables are $x^{(k)}j = \sum_{i=(j-1)k+1}^{jk} x_i$ and $\tau^{(k)}j = \sum_{i=(j-1)k+1}^{jk} \tau_i$. 
While the standard approach explores correlations between block variables, we focus instead on the statistics of rare events. 
Specifically, we examine the large $x$ behavior of $f^{(k)}(x)$ and the small $\tau$ behavior of $\psi^{(k)}(\tau)$, where $f^{(k)}(x)$ and $\psi^{(k)}(\tau)$ are the PDFs of the spatial and temporal block variables, respectively, after $k$ iterations of coarse-graining. 
When the rare event statistics of these PDFs transform as a function of $k$ in a way similar to IID RVs, we can treat the coarse-grained block variables as if they are IID for rare events.

Indeed, in a process with $N$ jumps, after $N$ coarse-graining iterations, the rare properties of $\sum_{i=1}^N x_i$ are captured by the large $x$ behavior of a single block variable, $x^{(N)}_1$. If, for $k > k^*$, the large $x$ limit of $f^{(k)}(x)$ transforms as a function of $k$ similarly to IID RVs, then the large $x$ properties of $f^{(N)}(x)$ will resemble those of IID variables with renormalized parameters. Once we know the large $x$ behavior of $f^{(N)}(x)$, we also know the large $X$ behavior of $P_N(X)$.
Similarly, since the large $N$ properties of $Q_t(N)$ are determined by the sum $\sum_{i=1}^N \tau_i$~\cite{wang2024statistics}, the same coarse-graining process applied to the temporal variables will yield $Q_t(N)$ in terms of renormalized IID RVs. By tracking how the $\tau \to 0$ limit of $\psi^{(k)}(\tau)$ transforms as a function of $k$, we can predict the behavior of $Q_t(N)$, provided that, for $k > k^*$, the transformations of $\psi^{(k)}(\tau)$ resemble those of the IID case, with renormalized parameters.



 We begin by deriving the forms of $f^{(k)}(x)$ and $\psi^{(k)}(\tau)$ for IID RVs.  For the spatial process, we assume that for large $|x|$, the PDF $f(x)$ follows: $f(x)\sim B |x|^{\alpha} e^{-\left(x/\delta\right)^\beta}$, while $\beta>1$. 
To obtain the behavior of the first coarse-grained PDF $f^{(1)}(x^{(1)}_1)$ for large $x^{(1)}_1 = x_1+x_2$, we compute the convolution integral: $\int_{-\infty}^\infty f(x_1)f(x^{(1)}_1-x_1){dx_1}$. 
For large $|x^{(1)}_1|$, the dominant contributions come from the large $|x|$ behavior of $f(x)$. Evaluating the integral gives: $f^{(1)}(x)\sim B^{(1)} {x}^{\alpha^{(1)}}e^{-(x/{\delta^{(1)}})^\beta}$, where the parameters transform as: $B^{(1)} = B^{2} \sqrt{\frac{2\pi\delta^\beta}{\beta(\beta-1)}} 2^{-\left(2\alpha-\frac{\beta}{2}+\frac{3}{2}\right)}$, $\alpha^{(1)} = 2\alpha + 1-\beta/2$ and $\delta^{(1)} = 2^{1-1/\beta}\delta$. 
By repeating the convolution process $k$ times, we obtain the coarse-grained PDF after 
$k$ iterations as:
\begin{equation}
    \label{eq:fkForm}
    f^{(k)}(x)\sim B^{(k)}|x|^{\alpha^{(k)}}e^{-(|x|/\delta^{(k)})^\beta}
\end{equation}
while the parameters evolve according to:
\begin{equation}
\label{eq:BTransform}
    \begin{array}{cl}
         B^{(k)}&=B^{k+1} \sqrt{\left[\frac{2\pi\delta^\beta}{\beta(\beta-1)}\right]^{k}} \left(k+1\right)^{-\left[(k+1)\alpha+k\left(1-\frac{\beta}{2} \right)+\frac{1}{2}\right]}  
         \\       
       \alpha^{(k)} &= \left(k+1\right)\alpha+k\left(1-\frac{\beta}{2} \right) 
       \\
        \delta^{(k)} &= \left(k+1\right)^{1-\frac{1}{\beta}} \delta 
    \end{array}
\end{equation}
For the temporal process, consider that for small $\tau$, the PDF $\psi(\tau)$ behaves as: $\psi(\tau)\sim C_0 \tau^{A}+C_1\tau^{A+1}$, where $C_0$, $C_1$, and $A$ are constants~\cite{barkai2020packets}. 
The first coarse-grained PDF $\psi^{(1)}(\tau_1^{(1)})$ is obtained via the convolution: $\psi^{(1)}(\tau^{(1)}_1)=\int_0^{\tau^{(1)}_1} \psi(\tau_1)\psi(\tau^{(1)}_1-\tau_1)d\tau_1$ in the $\tau^{(1)}_1\to 0$ limit. 
After $k$ iterations, the coarse-grained PDF is given by:
\begin{equation}
    \label{eq:psikForm}
    \psi^{(k)}(\tau)\sim C^{(k)}_0 \tau^{A^{(k)}}+C^{(k)}_1 \tau^{{A^{(k)}}+1}
\end{equation}
with parameters evolving as: 
\begin{equation}
  \label{eq:ATransform}
    \begin{array}{ll}
        A^{(k)}&=A+k(A+1) \\
        C^{(k)}_0 &= \left(C_0 A!\right)^{k+1}\big/\left([k+1][A+1]-1\right)! \\
          C^{(k)}_1 &= (k+1)C_1 \left(C_0 A!\right)^{k} (A+1)!\big/ \left([k+1][A+1]\right)!
    \end{array}
\end{equation}
Equations~\eqref{eq:fkForm}-\eqref{eq:ATransform} describe how PDFs for spatial and temporal processes evolve under repeated coarse-graining in the IID case. 
For correlated  processes, if the coarse-grained PDFs $f^{(k)}(x)$ and $\psi^{(k)}(\tau)$ evolve as a function of $k$ (for large enough $k$) similar to Eqs.~\eqref{eq:fkForm}-\eqref{eq:ATransform},  with effective parameters $B^e, \alpha^e, \delta^e, A^e, C_0^e, C_1^e$ instead of $B, \alpha, \delta, A, C_0, C_1$, then the emergence of Laplace tails is ensured.
These effective parameters will define the Laplace tails of Eq.~\eqref{eq:pxtLaplaceorigin} for correlated processes.
Next, we present two examples illustrating the emergence of Laplace tails for temporal and spatial correlations.

\begin{figure}[t]
 \centering
 \includegraphics[width=0.95\linewidth]{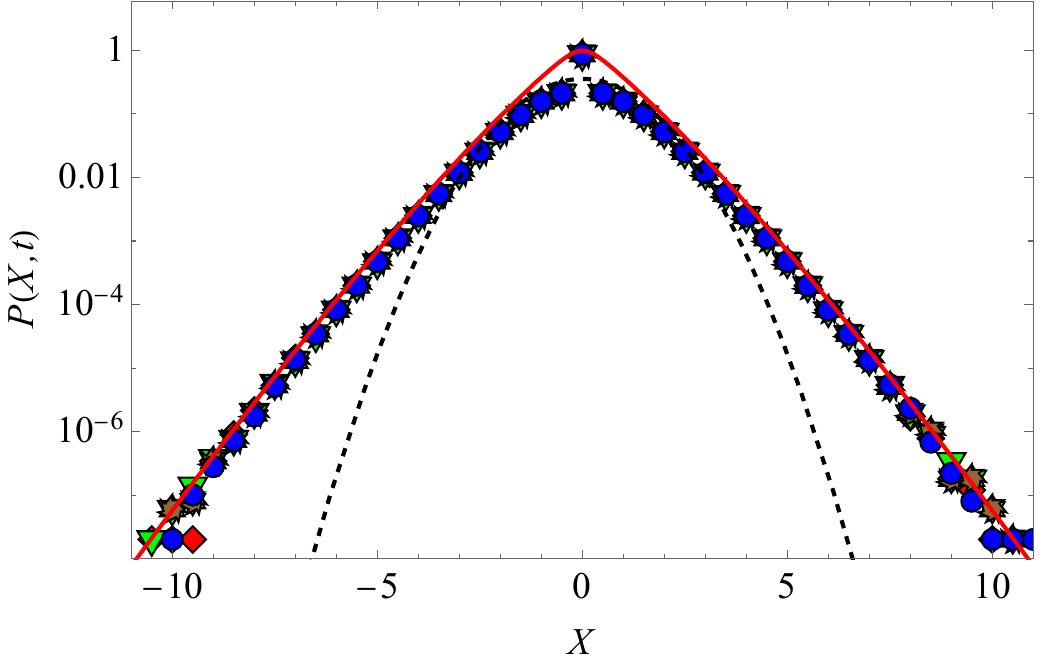}  
 \caption{
 $P(X,t)$ for a process with temporal correlations. Symbols represent numerical simulations, and the thick line corresponds to the theoretical result, while the dashed line shows the Gaussian approximation. For all values of the correlation parameter $\gamma$: $\gamma=0.5$ ({\color{green}$\nabla$}), $\gamma=0.25$ ({\color{blue}$\bigcirc$}), $\gamma=0$ ({\color{red}$\diamondsuit$}), $\gamma=-0.25$ (ratchet wheel), and $\gamma=-0.5$ ({\color{yellow}$\ast$}), the tails are identical. This is consistent with the theoretical result that the tails behave similarly to an independent process. The full theoretical form of $P(X,t)$ is provided in the SM. The other process parameters are $\omega = 1$, $\sigma = 1$, and measurement time $t = 1$. 
}\label{fig:temporalGauss}
\end{figure}

Example 1: Correlations in the temporal process. In this example, we introduce correlations in the temporal process, where each waiting time $\tau_i$ depends on the previous waiting time $\tau_{i-1}$. Specifically, we model the waiting times using a discrete version of the modified Ornstein-Uhlenbeck process for positive waiting times, see Supplemental Material (SM). The conditional  PDF of $\tau_i$ given $\tau_{i-1}$ is: $\psi(\tau_i|\tau_{i-1}) = \theta(\tau_i)\left(e^{-(\tau_i-\gamma\tau_{i-1})^2/2\omega^2}+e^{-(\tau_i+\gamma\tau_{i-1})^2/2\omega^2}\right)/\sqrt{2\pi\omega^2}$ where $\theta()$ is the Heaviside $\theta$ function, $\omega>0$ defines the intrinsic timescale, and $-1<\gamma<1$ is the correlation parameter.  
We set $\tau_0=0$. 
For $\gamma=0$ there are no correlations and the temporal process is renewal.
For the spatial process, we assume no correlations, and the jumps are IID Gaussian RVs, i.e., $f(x)=e^{-x^2/2\sigma^2}/\sqrt{2\pi\sigma^2}$, where $\sigma$ is the intrinsic length scale. 
Since the jumps are IID, the coarse-grained form $f^{(k)}(x)$ follows Eqs.~\eqref{eq:fkForm}-\eqref{eq:BTransform} and the effective parameters in the large $x$ limit are: $B^{e}=1/\sqrt{2\pi\sigma^2}$,$\alpha^{e}=0$, $\delta^e=\sqrt{2}\sigma$, while $\beta=2$. 
For the temporal process, $\psi^{(k)}(\tau)$ is the PDF of the sum of $k+1$ waiting times, i.e., $\psi^{(k)}(\tau)=\int_0^\tau\psi(\tau_1|\tau_0)\int_0^{\tau-\tau_1}\psi(\tau_2|\tau_2)\dots\int_0^{\tau-\sum_{i=1}^{k-1}}\psi(\tau_k|\tau_{k-1})\psi(\tau-\sum_{i=1}^{k}\tau_i|\tau_k)d\tau_k\dots\tau_1$. 
In the limit $\tau\to 0$ all  $\tau_i$  are small, and expanding $\psi(\tau_i|\tau_{i-1})$ around $\tau_i \to 0$ and $\tau_{i-1} \to 0$ gives: $\psi^{(k)}(\tau)\sim \frac{\left(\sqrt{2/\pi}/\omega\right)^{k+1}}{k!} \tau^{k} +O\left(\tau^{k+1}\right)$. 
Comparing this to Eqs.~(\ref{eq:psikForm}-\ref{eq:ATransform}) shows that in the limit $\tau \to 0$, coarse-graining of the temporal process matches that of IID RVs. The effective parameters are $A^e = 0$, $C_0^e = \sqrt{2}/\omega$, and $C_1^e = 0$. Importantly, the independence of $\psi^{(k)}(\tau)$ from the correlation parameter $\gamma$ in the $\tau \to 0$ limit indicates that the correlations between waiting times do not affect the large $|X|$ tails of $P(X,t)$.
Knowing the effective parameters $B^e$, $\alpha^e$, $\delta^e$, $A^e$, $C_0^e$, and $C_1^e$ allows us to apply standard methods (see Appendix and \cite{barkai2020packets,wang2020large,hamdi2024laplace}) for IID RVs to derive $P(X,t)$, which follows a similar form to Eq.~\eqref{eq:pxtLaplaceorigin}. The explicit form of the slowly varying function $Z()$ is given in SM. 
In Fig.~\ref{fig:temporalGauss}, we present results from numerical simulations for various values of $\gamma$ and compare them with theoretical predictions. The excellent agreement between simulations and theory, and the apparent independence from $\gamma$, support the conclusion that correlations in the temporal process do not affect the behavior of the tails in this example. 

\begin{figure}[t]
 \centering
 \includegraphics[width=0.99\linewidth]{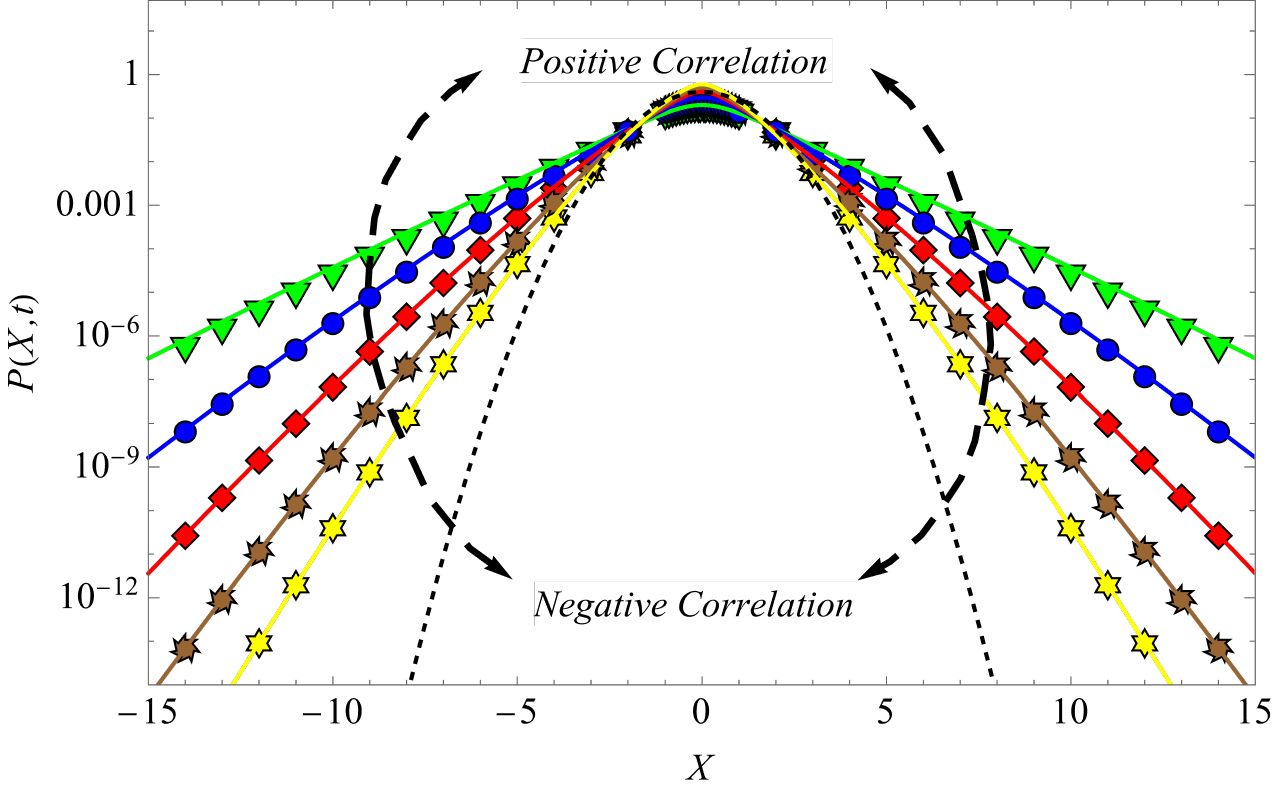}  
 \caption{$P(X,t)$ for the case of spatial correlations. Symbols represent numerical simulations, while thick lines correspond to the theoretical results. The dashed line shows the Gaussian approximation. The Laplace tails become more pronounced as the correlation parameter increases: $\gamma=0.5$ ({\color{green}$\nabla$}), $\gamma=0.25$ ({\color{blue}$\bigcirc$}), and decay faster as $\gamma$ decreases: $\gamma=-0.25$ (ratchet wheel), $\gamma=-0.5$ ({\color{yellow}$\ast$}), compared to the uncorrelated case $\gamma=0$ ({\color{red}$\diamondsuit$}). The full theoretical form of $P(X,t)$ is provided in SM. The other process parameters are $\omega=1$, $\sigma=1$, and measurement time $t=1$.
}\label{fig:spatialGaus}
\end{figure}

Example 2: Correlated spatial process. 
The conditional PDF of jump $x_i$ given the previous jump $x_{i-1}$ is modeled using a discrete modified Ornstein-Uhlenbeck process (see SM): $f(x_i|x_{i-1})=e^{-\left(x_i-\gamma x_{i-1}\right)^2/2\sigma^2}/\sqrt{2\pi\sigma^2}$. Here, $\sigma$ is the intrinsic length scale, and $-1 < \gamma < 1$ is the correlation parameter. 
The temporal process is independent with waiting times following an exponential distribution, $\psi(\tau_i)=\exp(-\tau_i/\omega)/\omega$, where $\omega$ is the intrinsic timescale. 
Positive correlations are present when $0 < \gamma < 1$, while negative correlations exist for $-1 < \gamma < 0$.
The temporal process has no correlations and the $\tau\to 0$ expansion of $\psi(\tau)$ provides the effective parameters:  $A^e=0$, $C^{e}_0=1/\omega$ and $C^{e}_1=-1/\omega^2$. 
The PDF of the coarse-grained spatial process, $f^{(k)}(x)$ is obtained by  summing  $x=\sum_{i=1}^{k+1} x_i$. 
Since $f(x_i|x_{i-1})$ is Gaussian, each $x_i$ can be represented as a weighted sum of standard IID Gaussian RVs $\tilde{x}_j$, where: $x_i=\sigma\sum_{j=1}^i \gamma^{i-j} {\tilde x}_j$. 
Summing over $k+1$ geometric series yields:  $x= \sigma\sum_{j=1}^{k+1} \frac{1-\gamma^{k+2-j}}{1-\gamma}{\tilde x}_j$. 
Since ${\tilde x}_j$ are Gaussian IID RVs, $x$ is also Gaussian and $\text{Var}(x)=\sigma^2\sum_{i=1}^{k+1} \left(\frac{1-\gamma^{k+2-j}}{1-\gamma}\right)^2\text{Var}({\tilde x}_j)$. Thus, the PDF $f^{(k)}(x)$ is Gaussian: $f^{(k)}(x)=e^{-\left(x/\delta^{(k)})\right)^2}/\sqrt{\pi\delta^{(k)}}$ where $\delta^{(k)}=\left(k+1\right)^{1/2}\frac{\sqrt{2}\sigma}{1-\gamma}\left(1-\frac{1}{k+1}\frac{\gamma(1-\gamma^{k+1})(2+\gamma[1-\gamma^{k+1}])}{1-\gamma^2}\right)^{1/2}$ while $B^{(k)}=1/\sqrt{\pi\delta^{(k)}}$, $\beta=2$ and $\alpha^{(k)}=0$. 
For large $k$, when the term $\left( 1 - 1/(k+1) \dots \right)^{1/2}$ converges to $1$, $f^{(k)}(x)$ behaves as in the IID scenario, with effective parameters: $\delta^{e}=\sqrt{2}\sigma/(1-\gamma)$, $B^{e}=1/\sqrt{2\pi\sigma^2/(1-\gamma)^2}$, and $\alpha^{e}=0$. 
Since we now have all the effective parameters, we can apply standard methods (SM and \cite{barkai2020packets,wang2020large,hamdi2024laplace}) to compute $P(X,t)$, which shows Laplace tails (Eq.~\eqref{eq:pxtLaplaceorigin}) in the large $|X|$ limit. The explicit form of the slowly varying function $Z()$ is provided in SM.
In Fig.~\ref{fig:spatialGaus}, , we compare the theoretical results with numerical simulations for different values of $\gamma$. The agreement is excellent, showing that positive correlations ($\gamma > 0$) lead to more pronounced tails, while negative correlations ($\gamma < 0$) result in faster decay. The case without correlations ($\gamma = 0$) is indicated by $\diamondsuit$ symbols in Fig.~\ref{fig:spatialGaus}.
As $\gamma \to 1$, the effective length scale $\delta^e = \sigma / (1 - \gamma)$ increases, leading to slower decay. Conversely, for $\gamma < 0$, the decay is faster. This behavior follows from Eq.~\eqref{eq:pxtLaplaceorigin}, where $\delta$ controls the inverse of the decay rate. Since $\delta^e$ acts as the effective $\delta$, a larger $\delta^e$ corresponds to a slower decay rate, and vice versa. Thus, as $\gamma$ increases, the decay rate decreases, leading to more pronounced tails, while negative $\gamma$ causes faster decay due to smaller $\delta^e$.
Additionally, as Eq.~\eqref{eq:starttail} shows, the effective parameters determine how close the tails are to the center, i.e., $X = 0$. A smaller $\delta^e$ produces narrower tails that start closer to the center, while a larger $\delta^e$ results in broader tails starting farther from the center.
Finally, for large $\delta^e$, many steps must be coarse-grained into a single block variable, which can delay convergence to IID behavior (see SM).

The general results and the examples of this work demonstrate that when rare events are concerned, correlations can be translated into effective parameters of a non-correlated process. 
As a result, the universality of Laplace tails is not restricted to IID RVs. Similar to Gaussian universality, Laplace tails emerge even when the microscopic dynamics are not independent.
However, unlike Gaussian behavior, which typically arises at long timescales, Laplace tails will appear at short times despite correlations, as dictated by Eq.~\eqref{eq:starttail}. 
These tails dominate when $|X|$ is large, and many steps must be taken to reach the distant location.
When correlations are not too extensive,  they mainly adjust the intrinsic time and length scales without altering the presence of Laplace tails at large $|X|$. 
Specifically, correlations determine the scale at which exponential decay will be detected and the rate of that decay. 

Notably, a process can exhibit correlations yet still behave as if uncorrelated for extreme events. This was illustrated in the first example.
This occurs because  Eqs.~\eqref{eq:fkForm}-\eqref{eq:ATransform} lead to IID-like behavior by focusing on the tails of the temporal and spatial PDFs rather than the average correlations. 
Equations ~\eqref{eq:fkForm}-\eqref{eq:ATransform}, not the specific correlation structure, provide a practical framework for deciding whether correlations will affect rare events and identify the effective parameters that describe rare events.

For granular materials, recent observations show that increased memory is associated with more pronounced exponential decay of the Van Hove function~\cite{yuan2024creep}. 
Our results, particularly the spatial correlations example, explain this phenomenon. The developed framework applies broadly to systems with inherent correlations~\cite{burov2011residence,majumdar2020toward,lanoiselee2018diffusion,wexler2020dynamics}. In active matter~\cite{bechingeractive,smith2023nonequilibrium}, such as transport driven by molecular motors~\cite{burov2013distribution,sonn2017scale}, where correlations and persistence play a crucial role, it provides a direct connection between correlations and the behavior of extreme events.

\begin{acknowledgments} {\bf Acknowledgments.} This work was supported by the  Israel Science Foundation Grant No. 2796/20.
\end{acknowledgments}

\bibliography{tails_corr.bib}

\end{document}


\setstcolor{red}
\title{ Supplemental Material for: The Emergence of Laplace Universality in Correlated Processes 
}

\author{R. K. Singh}
\affiliation{Department of Physics, Bar-Ilan University, Ramat-Gan 5290002,
Israel}

\author{Stanislav Burov}
\affiliation{Department of Physics, Bar-Ilan University, Ramat-Gan 5290002,
Israel}

\begin{abstract}
In this supplemental material, we present: {\bf I}. A detailed description of the discrete Ornstein-Uhlenbeck process, {\bf II}. The explicit form of $P(X,t)$ for a process with temporal correlations, and {\bf III}. The explicit form of $P(X,t)$ for a process with spatial correlations.  
\end{abstract}

\maketitle

\renewcommand{\theequation}{S\arabic{equation}}

\section{Discrete Ornstein-Uhlenbeck Process}

The Ornstein-Uhlenbeck process is governed by the Langevin equation: 
\begin{equation}
    \label{eq:oulangdef}
    dx_t = -b x_t dt+{\tilde\sigma} dW_t
\end{equation}
where $W_t$ represents the Wiener process, and the parameters satisfy ${\tilde\sigma}>0$ and $b>0$. This equation describes a particle under the influence of a harmonic potential, driven by white noise. The associated Fokker-Planck equation governs the evolution of the conditional probability density function (PDF), $P(x,t|x',t')$, representing the probability of observing $x$ at time $t$ given that the value was $x'$ at time $t'$: 
\begin{equation}
    \label{eq:fokkerplanck}
    \frac{\partial P(x,t|x',t')}{\partial t} = b \frac{\partial}{\partial x} \left(xP(x,t|x',t')\right)+\frac{{\tilde\sigma}^2}{2}\frac{\partial^2}{\partial x^2} P(x,t|x',t')
\end{equation}
The solution to Eq.\eqref{eq:fokkerplanck} is given by:
\begin{equation}
    \label{eq:pxtOUcont}
    P(x,t|x',t') = \frac{1}{\sqrt{2\pi {\tilde\sigma}^2 \left(\frac{1-e^{-b(t-t')}}{2b}\right)}}
    e^{-\frac{\left(x-e^{-b (t-t')}x'\right)^2}{2{\tilde\sigma}^2\left(\frac{1-e^{-b(t-t')}}{2b}\right)}}
\end{equation}
 which holds for continuous $x$ and $t$. To transition to discrete time, we set $t - t' = 1$, introducing the parameters $\gamma = \exp(-b)$ and ${ \sigma} = {\tilde\sigma} \sqrt{\frac{1-e^{-b}}{2b}}$. This leads to the conditional PDF:  
\begin{equation}
    \label{eq:pxtoudiscrete}
    P(x|x')=\frac{1}{\sqrt{2\pi{\sigma}^2}}e^{-\frac{\left(x-\gamma x'\right)^2}{2{\sigma}^2}}
\end{equation}
which describes the probability of observing $x$ given that the previous value was $x'$. While $\gamma$ and ${\sigma}$ are both dependent on the parameter $b$, we treat them as independent for the sake of generality. Thus, for the discrete Ornstein-Uhlenbeck process, the conditional PDF is given by Eq.\eqref{eq:pxtoudiscrete} with parameters $-1 < \gamma < 1$ and ${\tilde \sigma} > 0$. The parameter $\gamma$ represents the correlation strength, determining how strongly the value $x'$ influences the PDF of $x$. For $\gamma > 0$, the correlation is positive, favoring values of $x$ near $\gamma x'$. When $\gamma = 0$, $x$ is independent of $x'$, while for $\gamma < 0$, the preferred value of $x$ is $-|\gamma| x'$.

In the main text, we employ Eq.~\eqref{eq:pxtoudiscrete} to describe spatial correlations. A series of values $x_1, x_2, x_3, \dots$, generated using the conditional PDF $P(x_i|x_{i-1})$, exhibits correlations. For $\gamma > 0$, the correlation $\langle x_i x_j \rangle$ decays exponentially with $|i - j|$, provided $\gamma < 1$. For $\gamma < 0$, the correlation function decays and oscillates as a function of $|i - j|$.

For temporal correlations, we use the same conditional PDF but replace $x$ with the absolute value of the waiting times $\tau = |x|$, ensuring positivity. The PDF for this case is given by: 
\begin{equation}
    \label{eq:pxtOUtempdiscrete}
    P(\tau|\tau')=
    \frac{\theta(\tau)}{\sqrt{2\pi{\sigma}^2}}\left(
    e^{-\frac{\left(\tau-\gamma \tau'\right)^2}{2{\sigma}^2}}
    +
    e^{-\frac{\left(\tau+\gamma\tau'\right)^2}{2{\sigma}^2}}
    \right)
\end{equation}
where $\theta(\tau)$ is the Heaviside function, ensuring the positivity of the temporal variable $\tau$. As with the spatial case, $\gamma$ governs the correlation strength. A sequence of waiting times $\tau_1, \tau_2, \tau_3, \dots$ generated from this conditional PDF exhibits temporal correlations based on the value of $\gamma$.

\section{$P(X,t)$ for process with temporal correlations}

We assume the effective parameters $B^e$, $\alpha^e$, $\delta^e$, $A^e$, $C_0^e$, and $C_1^e$ are known and use the subordination method to find $P(X,t)$: 
\begin{equation}
    \label{eq:subordiantion}
    P(X,t) = \sum_{N=0}^\infty Q_t(N) {\cal P}_N(X).
\end{equation}
where ${\cal P}_N(X)$ is the probability of observing $X$ after exactly $N$ jumps. Since the spatial process is uncorrelated and $f(x)$ is Gaussian, we have:
\begin{equation}
    \label{eq:pnxGauss}
    {\cal P}_N(X)=\frac{1}{\sqrt{\pi N\left({\delta^e}\right)^2}}e^{-N \left(\frac{X}{N\delta^e}\right)^2}.
\end{equation}
The function $Q_t(N)$ represents the probability of observing $N$ jumps by time $t$. In the large $N$ limit, we use the result from \cite{barkai2020packets}: 
\begin{equation}
    \label{eq:qtnEq}
    Q_t(N)\underset{N\to\infty}{\sim}
    \frac{\left(\left\{C_0^e\Gamma[A^e+1]\right\}^{\frac{1}{A^e+1}}t\right)^{N(A^e+1)}}{\Gamma[N(A^e+1)+1]}e^{\left(C_{1}^e/C_0^e\right)t},
\end{equation}
where $\Gamma[\dots]$ is the $\Gamma$-function.
In the large $X$ limit, only summands with large $N$ contribute to the sum in Eq.~\eqref{eq:subordiantion}~\cite{barkai2020packets}.
Substitution into Eq.~\eqref{eq:subordiantion} of Eq.~\eqref{eq:pnxGauss}, Eq.~\eqref{eq:qtnEq} and the parameter values from Example 1 of the main text, leads to 
\begin{equation}
    \label{eq:pxtsumm01}
    P(X,t)\underset{|X|\to\infty}{\sim} \sum_{N=0}^\infty 
    \frac{\left(\sqrt{\frac{2}{\pi}}\frac{t}{\omega}\right)^N}{N!}
    \frac{1}{\sqrt{2\pi N\sigma^2}}e^{-X^2/2N\sigma^2}
\end{equation}
Approximating the sum as an integral gives:
\begin{equation}
    \label{eq:pxtint01}
    P(X,t)\underset{|X|\to\infty}{\sim} \int_0^\infty 
    \frac{1}{2\pi N \sigma} e^{-K(N)}\,dN
\end{equation}
where: 
\begin{equation}
    \label{eq:kndef}
       K(N)=\frac{X^2}{2N\sigma^2} - N \ln\left(\sqrt{\frac{2}{\pi}}\frac{t}{\omega}\right)+N\ln\left(N\right)-N
\end{equation}
where we used Stirling's approximation for $N!$. 
We find that $N^*$ for which $\frac{dK(N)}{dN}|_{N^*}=0$ satisfies 
\begin{equation}
    \label{eq:nstar}
    N^*=\frac{|X|}{\sigma} \big/\sqrt{W_0\left[\left(\frac{\omega X}{\sigma \sqrt{2/\pi} t}\right)^2\right]},
\end{equation}
and $W_0[z]$ is the principal branch of the Lambert $W$ function that for large values of $z$ has the asymptotic form $W_0[z]\sim \ln(z)-\ln\left(\ln(z)\right)$. 
Next, we use the second order expansion of $K(N)$, i.e., $K(N^*)+\frac{1}{2}\frac{d^2K(N)}{dN^2}|_{N^*}\left(N-N^*\right)^2$ and finally obtain
\begin{widetext}
    \begin{equation}
        \label{eq:pxttempfullform}
        P(X,t)\underset{|X|\to\infty}{\sim}
        \frac{1}{\sqrt{2\pi \sigma |X|\left(W_0\left[\left(\frac{\omega X}{\sigma \sqrt{2/\pi} t}\right)^2\right]^{\frac{1}{2}}+W_0\left[\left(\frac{\omega X}{\sigma \sqrt{2/\pi} t}\right)^2\right]^{-\frac{1}{2}}\right)}}
        e^{-\frac{|X|}{\sigma}
        \left(W_0\left[\left(\frac{\omega X}{\sigma \sqrt{2/\pi} t}\right)^2\right]^{\frac{1}{2}}-W_0\left[\left(\frac{\omega X}{\sigma \sqrt{2/\pi} t}\right)^2\right]^{-\frac{1}{2}}\right)}
    \end{equation}
\end{widetext}
Comparison of Eq.~\eqref{eq:pxttempfullform} with Eq.~(2) of the main text provides the form of the slow-varying function $Z(y)$:
\begin{equation}
    \label{eq:funcZtemp}
    Z(y) = W_0\left[\left(\frac{\omega }{\sigma \sqrt{2/\pi} }y\right)^2\right]^{\frac{1}{2}}-W_0\left[\left(\frac{\omega }{\sigma \sqrt{2/\pi} }y\right)^2\right]^{-\frac{1}{2}}
\end{equation}
which introduces the logarithmic correction to the exponential decay.  
The theoretical form of $P(X,t)$ in Fig. 1 of the main text is plotted using Eq.~\eqref{eq:pxttempfullform}.
The independence of $\omega$ and $\gamma$ ensures that the tails of $P(X,t)$ are unaffected by the correlation $\gamma$ (see Fig. 1 in the main text). However, in the special case where $\gamma$ and $\omega$ depend on $b$, the effective parameter $C_0^e$ is influenced, leading to modifications in the tails. 

\section{$P(X,t)$ for process with spatial correlations}

The effective parameters for the correlated spatial process are given in the main text. By repeating the derivation from the previous section, we find that $P(X,t)$ for the spatially correlated Ornstein-Uhlenbeck process takes the form of Eq.~\eqref{eq:pxttempfullform}, multiplied by $\exp(-t/\omega)$, with replacements $\omega/\sqrt{2/\pi} \to \omega$ and $\sigma \to \sigma/(1-\gamma)$. Setting $X/t = y$ and taking the large $|X|$ limit, we obtain the large deviation rate function:
\begin{equation}
    \label{eq:largedevrate}
    \begin{array}{l}
    \lim_{|X|\to\infty}\left(-\frac{1}{|X|}
    P(X,t)\right)=
    \\\frac{1-\gamma}{\sigma}
    \left(
W_0\left[\left(\frac{(1-\gamma)\omega }{\sigma }y\right)^2\right]^{\frac{1}{2}}-W_0\left[\left(\frac{(1-\gamma)\omega }{\sigma }y\right)^2\right]^{-\frac{1}{2}}\right)+\frac{y^{-1}}{\omega }.
\end{array}
\end{equation}

Eq.~\eqref{eq:largedevrate} defines the slow-varying function $Z(y)$ for large $|X|$, with corrections potentially arising for finite $X$ due to power-law convergence to the effective parameters. In the main text, we derived $f^{(k)}(x)$ and found $\delta^{(k)}$, with the limit $\delta^e = \sqrt{2}\sigma/(1-\gamma)$. If corrections are considered (neglecting powers of $\gamma^{k+1}$), the effective parameter $\delta^e$ depends on $N$ as:
\begin{equation}
    \label{eq:deltakEff}
    \delta^e=\sqrt{2}\sigma/(1-\gamma)\left(1-\frac{1}{N}\frac{\gamma(2+\gamma)}{1-\gamma^2}\right)^{1/2}
\end{equation}
In this case, $P(X,t)$ follows the integral form in Eq.~\eqref{eq:pxtint01} with:
\begin{equation}
    \label{eq:knspatial}
    \begin{array}{ll}
     K(N)= & \frac{(1-\gamma)^2X^2}{2N\sigma^2} - N \ln\left(\frac{t}{\omega}\right)+N\ln\left(N\right)-N
     \\
     &
     +\frac{(1-\gamma)^2 \gamma(2+\gamma)}{2\sigma^2(1-\gamma^2)} 
     \left(\frac{X}{N}\right)^2
     \end{array}
\end{equation}
The last term accounts for the power-law corrections. Using the saddle-point approximation without this correction term, the value of $N^*$ is
\begin{equation}
    \label{eq:nstarspatial}
    N^*=\frac{(1-\gamma)|X|}{\sigma} \big/\sqrt{W_0\left[\left(\frac{\omega (1-\gamma) X}{\sigma  t}\right)^2\right]},
\end{equation}
The correction term $(X/N)^2$ scales as $W_0\left[\left(\omega (1-\gamma) X/\sigma t\right)^2\right]$ and becomes increasingly negligible compared to the other terms in $K(N)$ as $|X|$ grows. However, for finite $X$, this term remains nonzero and must be considered when comparing simulation results to analytical predictions for finite $X$ and $t$. The corrected form of $P(X,t)$ for spatial correlations is:
\begin{widetext}
    \begin{equation}
        \label{eq:pxtspatialfullform}
        P(X,t)\underset{|X|\to\infty}{\sim}
        \frac{e^{-\frac{(1-\gamma)|X|}{\sigma}
        \left(W_0\left[\left(\frac{\omega (1-\gamma)X}{\sigma  t}\right)^2\right]^{\frac{1}{2}}-W_0\left[\left(\frac{\omega (1-\gamma) X}{\sigma  t}\right)^2\right]^{-\frac{1}{2}}\right)-\frac{t}{\omega}}}{\sqrt{\frac{2\pi \sigma |X|}{1-\gamma}\left(W_0\left[\left(\frac{\omega (1-\gamma) X}{\sigma  t}\right)^2\right]^{\frac{1}{2}}+W_0\left[\left(\frac{\omega (1-\gamma)X}{\sigma  t}\right)^2\right]^{-\frac{1}{2}}\right)}}
        e^{-\frac{\gamma(2+\gamma)}{2(1-\gamma^2)}W_0\left[\left(\frac{\omega (1-\gamma)X}{\sigma  t}\right)^2\right]}
    \end{equation}
\end{widetext}
The theoretical form of $P(X,t)$ in Fig. 2 of the main text is plotted using Eq.~\eqref{eq:pxtspatialfullform}.

\bibliography{tails_corr.bib}